%% file: nature_water.tex
\documentclass[11pt]{article}

\usepackage[margin=1in]{geometry}
\usepackage{graphicx}
\usepackage{booktabs}
\usepackage{amsmath}
\usepackage{amssymb}
\usepackage[numbers,sort&compress]{natbib}
\usepackage{setspace}
\usepackage[colorlinks=true,allcolors=black]{hyperref}

\newcommand{\ci}[2]{[#1, #2]}
\newcommand{\cif}[2]{[\mathord{\le}#1, #2]}

\title{\textbf{Don't look now: paying for documented need punishes
discovery in US lead-pipe funding}}

\author{Gustavo Pedro Ricou\thanks{School of Computer Science and Statistics,
Trinity College Dublin, College Green, Dublin D02 PN40, Ireland.
Correspondence: \texttt{pedrorig@tcd.ie}} \and
Kristin C. Epstein\thanks{CDM Smith, Edison, New Jersey 08837, United
States.}}

\date{}

\begin{document}
\maketitle
\onehalfspacing

\begin{abstract}
The United States allots \$15 billion for lead service line
replacement in proportion to documented lead, the largest of a growing
class of environmental-indicator transfers.
Reconstructed exactly, the rule prices looking below zero.
Resolving an unknown pipe lowers a state's expected allotment in three
of the eight jurisdictions we can measure above the statutory floor,
and raises it in none. Below it the allotment cannot move. Systems filing
nothing are scored lead-free. We prove no fixed formula on self-reported
inventories can pay for need without punishing its measurement, so we
made an audited one. The unique additive repair credits unknowns at a
rate no filing can move, and is strategy-proof. Discovery is
repriced from a median loss of \$48.63 per line per year to zero.
Allocation swings fall from 16.5\% of the pool to 0.8\%, for an audit
costing 1.23\% of annual funds. The formula's tilt toward vulnerable
systems survives.
\end{abstract}

\medskip

Removing lead pipes is among the largest public-health infrastructure
efforts the United States has attempted: \$15 billion over five years,
aimed at millions of service lines, divided among states by a federal
formula whose input is \emph{documented} lead.\cite{Humphreys2023} The Act
funding it is fifty years old and, on this journal's own reading, due
improvement.\cite{NatWater2024editorial,Hall2024} It is also the largest instance of a design
spreading worldwide.
Ecological fiscal transfers --- public revenue moved between governments
on environmental indicators --- grew from \$0.35 billion to \$23 billion
a year between 2007 and 2020, and the literature cataloguing them holds
that they ``in principle can incentivize'' the outcomes they
index.\cite{Busch2021} Whether they do depends on a quantity asserted
far more often than derived: the marginal price the formula puts on
\emph{measuring} the indicator.

Here we derive it, for the instrument where it can be computed exactly.
The Environmental
Policy Innovation Center showed by scenario comparison that EPA's lead
allotments advantage states with incomplete inventories, and could not
reproduce the rule itself.\cite{EPIC2024deepdive} We recover it as an exact
identity from EPA's memoranda and published tables, which lets
the marginal question --- what does a state gain or lose by investigating
one more unknown pipe? --- be asked at all. The answer is that the
formula pays states not to look. This is not a repairable detail
but a theorem: no fixed formula on a self-reported inventory can be
neutral to measurement and targeted at need at once. The repair must
enlarge the instrument class, and is unique within it among rules
additive across sub-populations.

\section{A formula that prices discovery below zero}

\begin{figure}[htbp]
  \centering
  \includegraphics[width=0.62\linewidth]{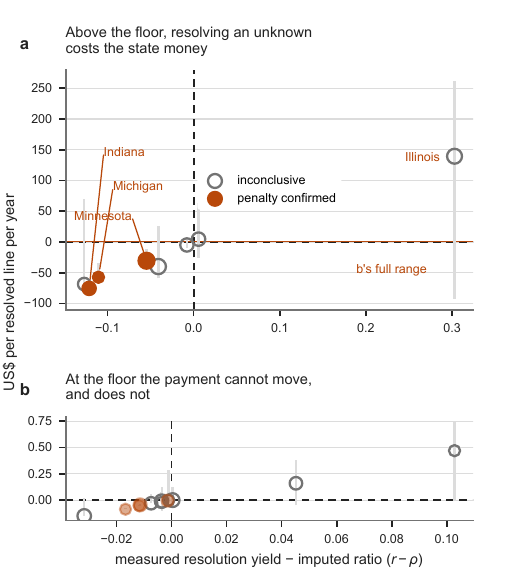}
  \caption{\textbf{The measured price of resolving an unknown line.}
  Expected federal dollars per resolved line per year against the gap
  between measured resolution yield $r$ and imputed ratio $\rho$. Of the
  21 states with enough resolution flow to measure $r$, the 20 whose
  records support an effect interval are shown, split by whether the
  payment can move. \textbf{a}, the eight above the statutory floor,
  where the derivative is not zero. \textbf{b}, the twelve measurable states at the floor, where
  the allotment is one per cent of the pool whatever the inventory says.
  Every effect there is inside \$0.5 a line a year against a median of
  \$48.63 in \textbf{a}, so the four confirmations in \textbf{b} are
  intervals around a quantity that cannot move: the marked band in
  \textbf{a} is the whole of \textbf{b}'s vertical range drawn to
  \textbf{a}'s scale, and is thinner than the rule beneath it. Both
  panels share the key in \textbf{a}. Vertical bars are 95\% intervals,
  cluster-bootstrapped over systems with 2,000 resamples except where
  Methods names the exact-binomial exception. Filled
  markers: penalty confirmed (interval
  excluding zero); open markers: inconclusive --- including
  Illinois at far right, whose large positive $r-\rho$ comes with an
  interval wide enough to span zero. No state is a confirmed reward
  under either definition of the resolution flow.}
  \label{fig:signlaw}
\end{figure}

Each state's projected lead is $P = L + \rho U$: documented lead and
galvanized lines $L$ at face value, unknown lines $U$ credited at the
state's classified lead ratio $\rho = L/(L+N)$, documented non-lead $N$
entering only through that ratio.\cite{EPA2025fy25} Shares of the pool
follow $P$, subject to a statutory floor of one percent.\cite{SDWA1452}
Resolving one unknown at true yield $r$ changes $P$, to first order, by

\begin{equation}
\mathbb{E}[\Delta P] \;\approx\; (r - \rho)\,\bigl(1 + U/C\bigr),
\qquad C = L + N,
\label{eq:signlaw}
\end{equation}

so the federal return to looking has the sign of $(r-\rho)$. The two
regimes also disagree about what an unknown \emph{is}: the compliance
rule counts one as lead for the replacement clock, while the formula
credits it at $\rho<1$. A utility recording honest uncertainty keeps
the whole obligation and is paid for part of it.\cite{LCRI2024} The level is
wrong too: $\rho$ is a synthetic estimator,\cite{Gonzalez1973,RaoMolina2015}
fitted to documented lines and applied to unknown ones, and systems
holding unknowns carry more documented lead --- so recomputing it from
the systems it prices moves \$60,566,911 between states (Supplementary
Note~2). In all
twenty-one states with enough resolution flow to measure $r$, the sign
predicted by equation~\eqref{eq:signlaw} matches the numerical derivative
of the full simulated pipeline --- a check on the reconstruction, not a test of behavior. With marginal cluster-bootstrap intervals, so the magnitudes carry the
claim rather than the count, the picture is one-sided
(Fig.~\ref{fig:signlaw}). A penalty is confirmed in three of the eight
we can measure, of nineteen above the floor
--- Indiana loses \$75.50 per resolved line per year
($\cif{-\$75.50}{-\$63.38}$), Michigan \$57.32, Minnesota \$30.51 --- and a
reward in none. At the floor the payment is
fixed and the derivative is zero: the median absolute effect there is
\$0.035 against \$48.63 above, a ratio of 1,389, and the four
confirmations among them measure a quantity that cannot move
(Fig.~\ref{fig:signlaw}b). Because the pool is fixed, every line one state
documents is financed by the other above-floor states, whatever the
national lead-line count.

\begin{figure}[htbp]
  \centering
  \includegraphics[width=\linewidth]{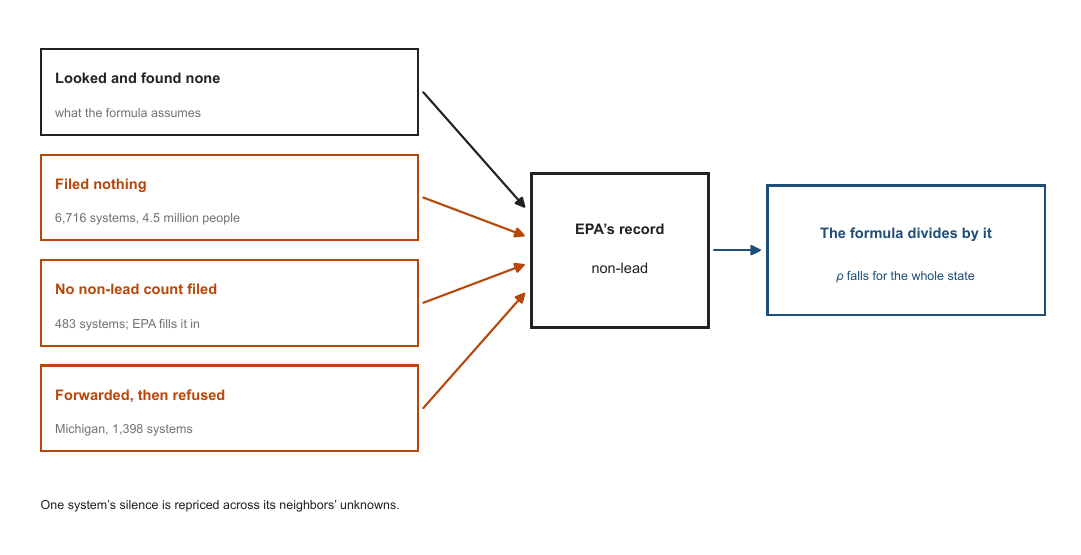}
  \caption{\textbf{Four ways to be recorded lead-free.} Only the first is
  a fact about pipes; the other three are facts about paperwork, and the
  allotment basis cannot tell them apart. Counts from the 2026Q2 export.
  Because a missing non-lead count is filled from a connection total, one
  system's silence lowers $\rho$ --- the crediting rate for every
  remaining unknown line in its state.}
  \label{fig:routes}
\end{figure}

The intensive margin sits on a harsher extensive one. A system that files
no inventory is scored as having no lead: 6,716 systems serving 4,535,279
people entered a \$2.7 billion allocation as lead-free. The count
the formula divides by --- non-lead lines --- was never a required federal
report, and EPA fills blanks by assuming them; 483 filers left it
blank.\cite{CFR14215,EPA2026allot} EPA's own Inspector General found
the questionnaire behind the FY2023 allotment carried minimal
verification.\cite{OIG2024} Filling a blank does not only zero
that system: it enlarges the state's non-lead denominator and lowers
$\rho$ for every remaining unknown. Silence is subsidized twice, and a filing can be
forwarded and refused federally, leaving a record identical to never
having filed (Fig.~\ref{fig:routes}).

Who bears this is not who a reader would guess. Systems serving more vulnerable populations hold more of their pipes
in the unknown column --- 9.7 percentage points more across the
vulnerability range, as a tract-level study of one city
found.\cite{Nigra2023} So the formula's weighting runs \emph{toward} vulnerability, by
\$13.42 per person, precisely because unknowns are credited at $\rho$
(Fig.~\ref{fig:equity}a). Zeroing that credit would fix the incentive by
moving money \emph{away} from vulnerable systems. The repair keeps the
credit and disciplines it: unknowns earn their expected lead at an
audited rate, and the tilt survives (Fig.~\ref{fig:equity}b).

Recipients price the output: seventeen states declined
\$871,406,603 of lead allotments in three
years.\cite{Phillis2023,EPA2025reallot} An allocation this brittle, and this signed,
invites redesign.

\section{Neutrality is a differential equation with no fixed solution}

Equation~\eqref{eq:signlaw} kept reappearing whichever pathology we
examined --- floor, silence subsidy and reclassification churn all reduce
to the same directional structure --- so neutrality is an equation to
solve rather than a property to test. A warehouse paid for counted stock stops counting shelves it thinks are
empty. Resolving $dq$ unknowns at true yield $r$ moves a state's
inventory along the direction $(dL, dN, dU) = (r,\, 1-r,\, -1)\,dq$.
Neutrality --- the payment unchanged by the act of measurement --- is
therefore a first-order partial differential equation on the payment
function $P(L,N,U)$:

\begin{equation}
r\,\frac{\partial P}{\partial L} + (1-r)\,\frac{\partial P}{\partial N}
- \frac{\partial P}{\partial U} = 0.
\label{eq:pde}
\end{equation}

Its characteristic curves conserve two quantities, the expected total lead
$L + rU$ and the expected total non-lead $N + (1-r)U$, so every neutral
payment is a function of them. Requiring the payment to target lead and to
be additive across sub-populations leaves the unique representative
$P^{*} = L + rU$. Additivity does work worth naming: it
excludes every concave or equity-weighted targeting rule --- equity is a
separate instrument, not a curvature inside this one. The current rule fails
equation~\eqref{eq:pde} by exactly the derivative in
equation~\eqref{eq:signlaw}, so the diagnosis and the design share one
arithmetic.

If
equation~\eqref{eq:pde} holds at just \emph{two} distinct yields,
subtracting the instances forces
$\partial P/\partial L = \partial P/\partial N = \partial P/\partial U$:
the payment sees total lines and cannot see lead at all. The agency does
not know $r$, so a fixed formula would need neutrality across a range of
yields --- and \textbf{no fixed formula on the filed inventory can be both
measurement-neutral and targeted}, an impossibility whose abstract form
is known.\cite{LiQiu2026} The target is not ours to choose:
EPA states its responsibility is to allot ``commensurate with the lead
service line replacement needs of each state''.\cite{EPA2026corrective} The credit rate must equal
the true yield, estimated somewhere the state's choices cannot reach:
fitted to the state's own resolutions, the derivative gains a term
$U\,d\hat r/dq$ --- how one utility's filing repriced Illinois
fourfold.\cite{Chicago2026} Proofs are in Supplementary Note~1.

The repair now on the table is inside the class the theorem closes. The
leading proposal would reweight unknown lines by inventory completeness and by a model of
how likely they are to be lead.\cite{EPIC2025bottlenecks}
Both are computed from the filed inventory, so the crediting rate
becomes a function of what the state files and the derivative returns
with its old sign. The instinct is right and the placement wrong:
predictive models are the right instrument for choosing which pipes to
investigate and the right prior for stratifying the audit that prices
them. They are the wrong rate to pay on, because the filing that improves the
model also moves the money.\cite{Hardt2016}

\section{An audit-anchored allocation}

The repair prices every line a state has, including the ones it never
mentions. With $K$ the state's service connections --- anchored in billing
records, not in the inventory priced --- and $M = K - (L+N+U)$ the lines
no filing covers, the need estimator is

\begin{equation}
\hat T \;=\; \hat v\,L \;+\; \hat r\,U \;+\; \hat\pi\,M,
\label{eq:estimator}
\end{equation}

Here $\hat v$, $\hat r$ and $\hat\pi$ are the raw, unshrunk means of
agency-drawn random audits of the three declared pools, once drawn. No
column is taken on trust, including the one the state fills in itself.
Only unbiasedness for each pool's mean is needed, so any frame the
primacy agencies already sample on will serve. Shrinkage would pay every
state partly for the mean of the others, and national rates survive only
as the cold start for a pool not yet audited.
Auditing filed inventories reuses the 2024 rule's validation machinery --- a
federal audit of state-reported data at this scale is an existing operation,
not a proposal.\cite{MACPAC2026perm} EPA now recommends resolving unknowns from a random sample behind a
model;\cite{EPA2026tips} sampling the lines no filing covers is a
further step under the Act's information-gathering
authority.\cite{SDWA1445} Decomposing the true lead $T$ over equation~\eqref{eq:estimator}'s
pools gives the identity
$\mathbb{E}[\hat T] = T - \ell_N$. The non-lead column is the only
unpriced pool, so hiding lead there lowers the payment and disclosing it
raises the payment back, while every other margin --- padding the lead
column, resolution order, filing or silence --- cancels identically. No
filing choice moves the estimate; audit noise moves the payment
(Methods), and silence is priced at prevalence rather than at zero. Found lead adds to $L$ what it removes
from the expected content of $U$, and the audit measures whatever pool
each choice leaves behind.

Auditing the front column buys the last of that. At face value it
leaves a term $(L - \ell_L)$ paying for over-declaration that no
penalty closes --- the statutory one is a ceiling per violation, the
gain per line.\cite{CFR19} The margin is not hypothetical: EPA de-obligated
\$314,429,000 from two states in 2026 after both over-claimed lead
lines, and GAO names distortion of formula inputs as a risk
class.\cite{EPA2026corrective,GAO2026fraud} Measuring the column removes the
term, for 5,840 lines and 0.11 points of the pool
(Supplementary Note~1).

\begin{figure}[htbp]
  \centering
  \includegraphics[width=0.62\linewidth]{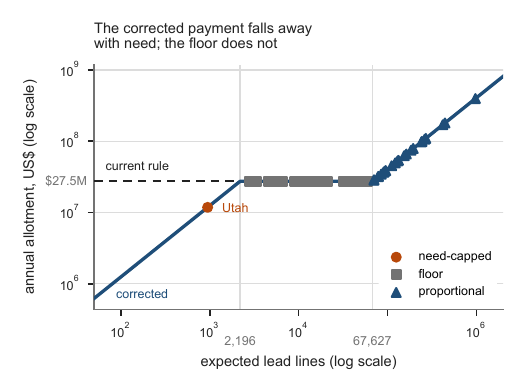}
  \caption{\textbf{The corrected allocation in closed form.} Payment
  against expected lead lines $\hat T$ under
  equation~\eqref{eq:allocator}, log--log, with states placed at the
  audited scenario's values and colored by regime. Each rule is named
  where it runs. A log axis cannot show $\hat T = 0$, so the atom is stated
  rather than plotted: the current rule pays \$27,456,000 there and
  equation~\eqref{eq:allocator} pays nothing, removing the atom by
  continuity.}
  \label{fig:mechanism}
\end{figure}

Four objectives get four instruments rather than one formula.
\textbf{Targeting}: allotments proportional to $\hat T$.
\textbf{Discovery}: a per-line bonus for validated resolutions,
funded from the \$129,375,000 annual gap between the appropriation and the funds available to states. With the pool neutral the return
to looking is positive --- the counterweight the compliance rule
does not supply until November 2027.\cite{Marcus2026} Each state draws in proportion to its unknown stock, retiring 1.1\% of
the national stock a year at \$524 a validated line, both sides. The
bonus pays for looking, not finding: paying on finds would reprice the
yield and undo the pool's neutrality.
\textbf{Small programs}: the statutory floor $F_0$ capped by need,
$f = \min(F_0,\, c\hat T)$ at replacement cost $c$.\cite{AWWA2022costs}
\textbf{Stability}: paid shares follow
$s \leftarrow s + \lambda(\hat s - s)$, with $\lambda$ a damping
weight estimated from the panel's own revisions, and audited baseline
resets bypassing the damper. The allocation itself then has a closed form,

\begin{equation}
\hat a(\hat T) \;=\; \max\!\bigl(\min(F_0,\, c\hat T),\; \theta^{*}\hat T\bigr),
\qquad \textstyle\sum_i \hat a(\hat T_i) = A,
\label{eq:allocator}
\end{equation}

continuous, nondecreasing, and --- unlike the current rule --- equal to
zero at zero need: the current allocation carries an atom of
\$27,456,000 at $\hat T = 0$, which equation~\eqref{eq:allocator}
removes by continuity rather than by punishment
(Fig.~\ref{fig:mechanism}). The regime boundaries are
intersections, not choices: need-capped below $F_0/c$ = 2,196 expected
lines, floored to $F_0/\theta^{*}$, proportional above. Share neutrality
follows from numerator neutrality, and the kinks add a curvature term
audit noise produces rather than any filing choice (Supplementary
Note~1).

\section{The instrument that revealed the defect calibrates its repair}

\begin{figure}[htbp]
  \centering
  \includegraphics[width=\linewidth]{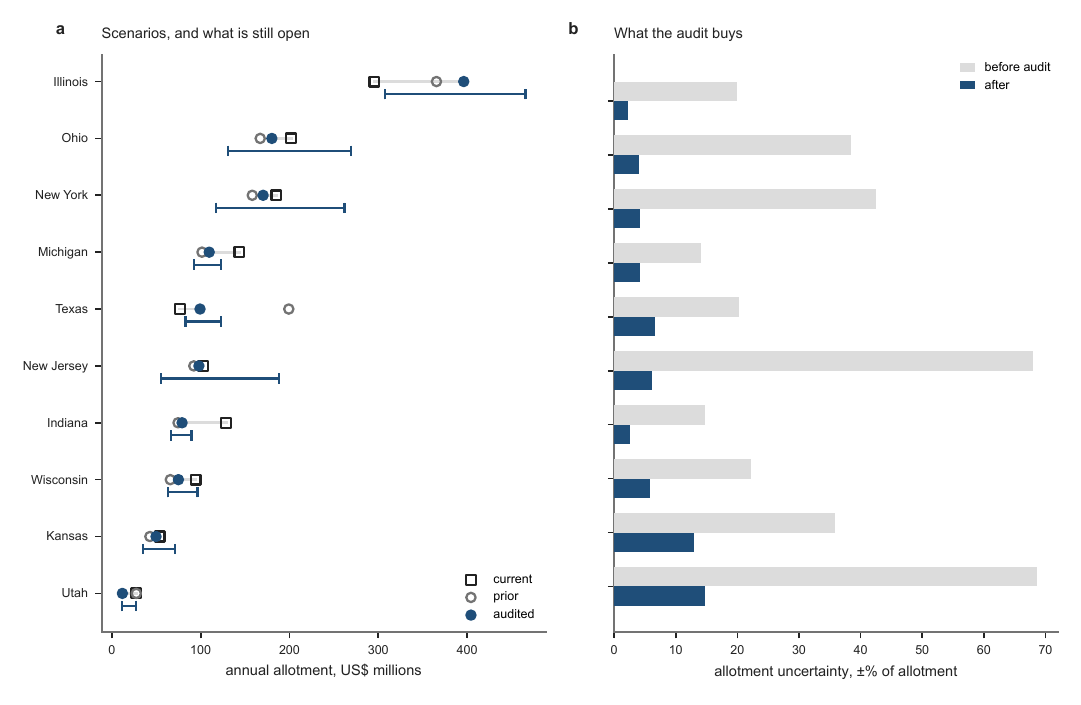}
  \caption{\textbf{What the repair moves, and what the audit still has to
  settle.} The ten largest movers between the two crediting scenarios.
  \textbf{a}, three marks a state: the rule in force (open square), the
  prior scenario (open circle) and the audited proposal (filled circle),
  with the connector spanning law to proposal and whiskers giving the
  pre-audit range --- each state's own yield drawn from the range its
  own measurements support, the audit drawn from the pool that results, and both
  carried through the whole allocation. Scenario and range are different
  estimands, so a scenario point may lie outside the range, as Texas's
  does. Against the rule in force the repair moves \$271 million, 10.0\%
  of the pool: 13 jurisdictions gain, 13 lose and 26 --- every one of
  them at the floor --- do not move at all. \textbf{b}, that range against
  the audit's own sampling range, as a percentage of the state's
  allotment; a written zero is a range the audit closes outright. Across
  these ten, today's evidence leaves the allotment open by $\pm$14\% to
  $\pm$69\%, and the audit closes that to $\pm$15\% or less. Utah is
  the exception that names the limit: its need cap turns on 48 anticipated
  lead lines in an unfiled pool of 53,644, which a sample meeting the
  national dollar tolerance still leaves open to $\pm$15\%. These are the
  states the scenarios move, not the states the evidence leaves most open
  (Supplementary Note~2). The distance
  between the bars is what the \$33.3 million both-sides audit buys;
  a records review, at \$1.5 million, reads the utility
  side and resolves nothing.}
  \label{fig:intervals}
\end{figure}

\begin{figure}[htbp]
  \centering
  \includegraphics[width=0.62\linewidth]{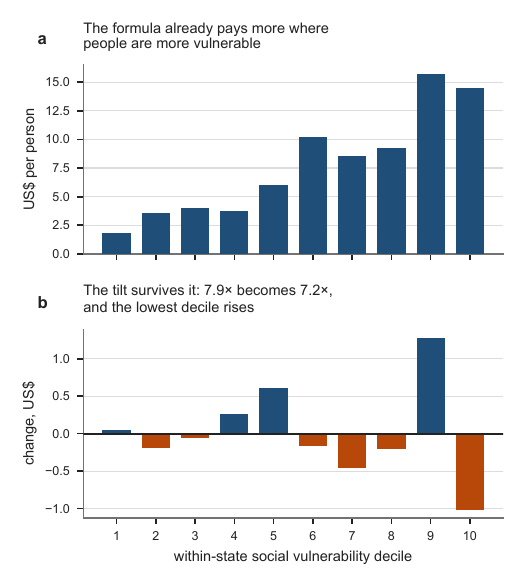}
  \caption{\textbf{Who the formula pays, and whether the repair changes
  it.} Federal dollars per person by within-state social vulnerability
  decile, population-weighted, across the 40,575 systems with an index
  value in a state that receives an allotment. \textbf{a}, the current
  rule already pays more where people are more vulnerable --- \$1.83 a
  person in the least vulnerable decile against \$14.47 in the most ---
  not by design but because those systems hold more of their pipes in
  the unknown column, and unknowns are credited at $\rho$. Zeroing that
  credit would fix the incentive by moving money away from them.
  \textbf{b}, what the audited estimator does to each decile. The tilt
  survives: the ratio between the extreme deciles goes from 7.9 to 7.2,
  the lowest decile rises rather than falls, and no decile moves by more
  than \$1.30. On the unweighted series the gradient widens slightly,
  from \$9.50 to \$9.76. The index is available for 65.5\% of reporting
  systems and 32.3\% of non-reporting ones; what that limits is set out
  in Supplementary Note~3.}
  \label{fig:equity}
\end{figure}

We compute the system on the panel that exposed the sign law --- six
quarterly vintages covering 66,414 water systems --- under two scenarios
bounding what the audit can find. \emph{Prior} credits silence and
unknowns at national rates, and \emph{audited}, crediting each at its own
measured rate (Table~\ref{tbl:calibration}). Four behaviors
demonstrate the mechanism. Texas, whose reported lines
cover 0.55 of its connections, halves between the two scenarios,
\$199.2 million to \$99.3 million (Supplementary Note~1).
Illinois, whose unknowns resolve to lead at the highest measured rate,
gains under both: the current formula paid it to look by accident, this
because the audit finds lead. Every confirmed penalty that can move
loses, and must: $r<\rho$ prices its unknowns above their audited
worth. And Utah is need-capped at \$11.8 million against a \$27.5 million
floor --- a saving of \$15.6 million at the average replacement cost,
nothing above \$29,028 a line (Supplementary Fig.~1). The evidence cannot settle it: no lead in 21,137 resolved lines, but across
only 35 systems the exact binomial bound reaches 8.2\%. The cap frees between nothing and
\$66.4 million (Fig.~\ref{fig:intervals}, Table~\ref{tbl:calibration}),
which argues for the audit, not against the cap: 626 lines
settle Utah for \$328,024. 

Every confirmed penalty is repriced to zero plus the bonus, and the
damper cuts the FY2025 swing from 16.5\% of the pool
to 0.8\%. Of the two interim priors only one survives auditing: halving
or doubling the yield moves 3.9\% to 6.6\% of the pool, while the
silence prior leaves the audited arithmetic entirely
(Supplementary Fig.~2).

The audit anchoring the system needs 63,510 lines nationally --- 28,106
in the unknown pools, 30,749 where no filing reaches and 4,655 verifying
declared lead --- at 1.23\% of one year's pool with both sides priced.\cite{CFR14190,EPA2026tips} Two thirds of the unknown-pool audit ends the cold start for 32
jurisdictions with no measurement of their own; the median state gives
107 lines. No share is a chosen
number. With $p^{*} = \varepsilon A/(\theta^{*} L)$ the padding fraction whose
payment error reaches the tolerance $\varepsilon A$ of \$2.7 million,
the declared-lead pool takes

\begin{equation}
m \;=\; \max\Bigl(\,\underbrace{m_{\min}}_{\text{floor}},\;
\underbrace{3/p^{*}}_{\text{detect}},\;
\underbrace{v(1-v)/p^{*2}}_{\text{precise}},\;
\underbrace{m_{R}}_{\text{regime}}\,\Bigr),
\qquad v = \tfrac{1}{2}.
\label{eq:sizing}
\end{equation}

\textbf{Detect}: with no exception found, $3/p^{*}$ rules out any rate
above $p^{*}$ at 95\% confidence, the standard for a population believed
clean.\cite{PCAOB2315} \textbf{Precise}: detection is not the promise. A sample that only detects breaks the tolerance once a declared
column is really 95\% lead --- the expected rate, not a pessimistic
tail. So the variance is bounded too, at its maximum $v = \tfrac12$,
the one evaluation point that is not itself a hypothesis about what is
measured (Supplementary Fig.~3). Every pool takes the floor, a
precision term and a \textbf{regime} criterion (Supplementary Note~1)
--- enough that the state's regime is not in doubt, which precision
alone misses because a pool holding almost no lead has almost no
variance to control. Detection is the declared column's alone.

$\hat v = 1$ is what the rule assumes, not what the column is worth:
a classification resting on a recent record or an inspection is accurate
to about 95\%, one resting on a predictive model from 50\%. Models clear
lines rather than declare them --- a statewide audit finds them
returning lead in no era where verification returns
1.5--14.5\%\cite{Sohail2026} --- so declared lead rests mostly on
records and inspection: 0.95 is the expectation and 0.50 a stress test
(Supplementary Note~4). Crediting it at the
expected 0.95 still relocates 0.37\% of the pool, \$10.0 million on an
assumption nothing tests; at the 0.50 stress test 5.3\%. Either way the
money moves \emph{toward} states whose need is mostly unknown and away
from the most confident inventories (Supplementary Fig.~4). Every parameter is estimated from the program's
own data or a stated convention with its consequence quantified
(Supplementary Table~1).

\section{Discussion}

The theorem is not about lead. Any transfer paying jurisdictions on a
self-reported indicator --- the \$23 billion a year the
ecological-fiscal-transfer literature catalogues --- faces
equation~\eqref{eq:pde}, whose impossibility says the incentive cannot
be fixed by reweighting inside the reported data. That literature's
``in principle can incentivize'' clause\cite{Busch2021} meets a derived
and measured counterexample, and the first indicator transfer proved
measurement-neutral. Payments are neutral exactly when the rate
crediting the unmeasured stock is set by an authority the recipient
cannot influence, and discovery-positive when a separate instrument pays
for the measurement.

The repair has a formal home. Allocation among agents with private
information about their own need, no transfers, and a costly
truth-revealing audit is a mature mechanism-design problem asking which
mechanism is optimal for a principal with an audit
budget.\cite{BenPorath2014,Mylovanov2017,Brzustowski2024} We ask which
are measurement-neutral, of a rule Congress already wrote, by the
differential route used since Rochet --- on the filed report rather than
a private type.\cite{Rochet1987,McAfee1988}

Equalization grants tax the measured tax base at the
margin;\cite{Smart1998,Buettner2006} performance systems distort the
dimension they score;\cite{HolmstromMilgrom1991} self-reported
indicators move the reporting.\cite{SandefurGlassman2015} Audits deter
poorly in the field and ours measures rather than
deters;\cite{KaplowShavell1994,DeLaO2023,Loch2024} index
insurance triggers on an index the insured cannot
move.\cite{Teh2019} And the class has measured pathologies of its own,
dilution in Brazil\cite{Ruggiero2022} and attenuation in
China.\cite{Xiao2025} What is new is not that an indicator
transfer distorts, but that the price the rule puts on
\emph{measurement} can be derived from the deployed formula as an identity
and confirmed in the payments it made. Dilution and attenuation are
congestion on the indicator; this is a negative price on resolving its
own missing data. EPA's August 2026 guidance encourages predictive models, and the record
shows they clear lines rather than declare them --- lowering $\rho$ and
the payment twice over.\cite{EPA2026tips} The federal incentive is prospective and the
test of whether states have already responded to it was uninformative.
The state-program incentive is neither --- it is operating now, and it
runs the other way.

FY2026 was the last IIJA lead round, which sharpens the question:
reallotments continue, the 2037 obligation stands, and Theorem~1
binds the successor rule.\cite{EPIC2026final}
The repair's demands are modest because its hardest components exist;
only the need cap needs a statutory amendment. A state that believes
its unknowns differ from the interim rate need not wait: the
discovery fund buys the audit that proves it. The
design cannot choose projects within states or replace the compliance
clock, and it reaches only the 52 jurisdictions the allotment does ---
leaving the tribal and territorial set-asides, whose own documented need
it never prices. Its guarantees are as strong as
the data beneath them, which is why the calibration is
scenario-bounded. Within those bounds the conclusion is plain: a formula can
pay for need or punish its discovery, but under the current design class
it cannot avoid both.

\begin{table}[htbp]
  \centering
  \caption{\textbf{Calibration on the FY2026 panel.} Need estimates and
  allotments under the two bounding scenarios, ordered by audited
  allotment; regimes from equation~\eqref{eq:allocator}. The bracket
  is the \emph{pre-audit} range: the 2.5th to 97.5th percentile of the
  audited allotment when each state's true yield is drawn from the range
  its own measurements support and the audit is then drawn from the pool
  that results, all carried through the whole allocation. It is what today's
  evidence leaves open, not what the audit's own sampling error does ---
  that is far narrower, and is the quantity the sizing rule makes a
  promise about. The full 52-jurisdiction table is Supplementary Table~2.}
  \label{tbl:calibration}
  \small
  \input{nw_table1}
\end{table}

\section{Methods}

\subsection{Data}

Six quarterly releases of EPA's national service line inventory
(2025Q1--2026Q2) covering 66,414 community and non-transient
non-community water systems; EPA's allotment memoranda and published
tables for FY2023--FY2026 --- the FY2026 appropriation was reduced by
\$125 million before allotment, so the \$2,704,441,000 pool and the
one per cent floor used here are post-rescission;\cite{EPA2026allot} service-connection counts from SDWIS; the
published reallotment memoranda recording declined
funds.\cite{EPA2025reallot,EPA2023reallot22} Five of the six quarterly
releases no longer exist upstream --- EPA overwrites its dashboard each
quarter --- so the panel cannot be rebuilt by anyone starting today. All
archived files carry checksummed provenance records verified in
continuous integration.

\textbf{Universe.} The inventory covers 66 primacy agencies; the
allotment reaches \textbf{52} --- the fifty states, the District of
Columbia and Puerto Rico --- so the allocation is computed over those
and no others. The fourteen outside are EPA direct-implementation
regions and the remaining territories, funded instead through the Act's
tribal and territorial set-asides.\cite{SDWA1452} They hold 1,168
systems serving 2,019,772 people, and between them \textbf{six} declared
lead lines: 0.000\% of the national total, so no quantity in this paper
turns on their exclusion. They do hold 1.9\% of declared galvanized
lines requiring replacement, almost all in the US Virgin Islands, and
0.5\% of unknowns, half of those in Guam --- documented need that the
formula analyzed here structurally never reaches, which is a fact about
the instrument rather than a limitation of the estimate. Nothing in the
design depends on the universe being these 52: the estimator and
equation~\eqref{eq:sizing} apply to any pool divided among any set of
recipients reporting the same way, the set-asides included.

\subsection{Reconstruction of the current rule}

The allotment rule is recovered as an exact identity, not a fitted line:
dividing each published grant by the published base reproduces every
printed percentage to both decimals across all 52 jurisdictions.
Equation~\eqref{eq:signlaw} is the directional derivative of the
recovered rule along the resolution direction; it was verified against
the numerical derivative of the full simulated pipeline in all 21 states
with sufficient resolution flow. Three counts differ and each is exact:
21 states have enough flow for the derivative to be computed, 20 of
those have records supporting an effect interval and are plotted, and 20
clear the estimability bar --- at least 20 systems and 5,000 resolved
lines --- to be credited at their own rate rather than at $r_0$. The two
twenties are not the same twenty: Iowa is plotted and not estimable. Resolution yields $r$ are net lead found
per resolved line from the quarterly panel, with
system-clustered intervals; reconciliation-class events are re-estimated
separately, and no state is a confirmed reward under either definition.
Two estimators produce those intervals, and which one applies is a
property of the data rather than a choice. Eighteen of the twenty are 95\%
cluster-bootstrap percentile intervals, 2,000 resamples, with systems as
clusters. For the
two states where no system recorded a positive find the bootstrap is
degenerate --- the yield is clipped at zero, so every replicate returns
the same value --- and the upper bound is instead the exact one-sided
binomial bound $1-\alpha^{1/k}$ on the $k$ systems, the rule of three at
$\alpha = 0.05$: Utah at $k = 35$ and Washington at $k = 112$. Ten states
have a yield sitting at the clip, and for those the interval is one-sided
by construction, so the reported loss is a floor rather than a center.
Details, including the treatment of the one state whose records are too
thin for an interval, follow the companion analysis in Supplementary
Note~2.

\subsection{What the floor costs, and that we keep it}

The repair reprices discovery from a median loss to zero, and zero is
where it stops for the twenty-seven jurisdictions the floor pays. Their
allotment is $F_0$ whatever their inventory says, so
$\partial a_i/\partial\hat T_i = 0$ exactly --- which is the property this
paper indicts in the rule in force, reproduced by the repair for 27 of 52
states and \$741 million, 27.4\% of the pool. It is not an oversight in
the repair. It is what a guarantee is: any minimum has zero slope below
the point the formula overtakes it.

It could be removed only by lowering the guarantee. Paying floor states
$F_0' + \theta^{*}\hat T_i$ with the block held at its present cost puts
$F_0'$ at \$19.0 million, 0.70\% of the pool against the statutory one per
cent. The need cap already asks Congress for less than the floor where a
state's need cannot absorb it; this would ask it for every low-need state,
which is a different and much larger request, and we do not make it. The
discovery bonus is the instrument that pays those states for looking, and
it is priced per validated line rather than through the allotment
(Supplementary Note~3) precisely because the allotment cannot carry it.

\subsection{Who draws the sample, and why it need not be the agency}

Theorem~2 requires each credited rate to be a quantity the recipient's own
choices cannot move. It is tempting to read that as requiring the primacy
agency to run the audit, and practitioners are right that it could not: an
agency has neither the crews nor the devices, and would have to rely on each
system to do the work and report. But the theorem does not need the agency
to hold a shovel. It needs the recipient not to choose the sample. Who digs
and who picks are separable, and separating them is what makes the design
operable.

The rule already supplies most of the apparatus.
\S141.84(b)(5) requires water systems to validate a random subset of
non-lead lines, to draw substitutes where access fails rather than accept a
short sample, and to submit results together with the specific version and
date of the inventory the pool was drawn from.\cite{EPA2024validation} Three
things are missing, and each has a cheap fix.

\textbf{The draw is run by the party being measured.} The rule permits "a
random number generator or lottery method" with no seed and no
reproducibility requirement, which is a lottery nobody outside the system
can check. The repair is a sequence rather than a technology, in three
steps.

First, the system publishes a \emph{checksum} of the list of lines it will
draw from: a short string of characters computed from the list by a
standard calculation, which comes out completely different if any line is
added, removed or altered, and from which the list itself cannot be
recovered. It is the tamper-evident seal on a sample bottle. The seal says
nothing about the contents and everything about whether they changed after
sealing. The versioned inventory \S141.84(b)(5) already requires is exactly
what it seals.

Second, and only once every system has filed its checksum, the agency
publishes a single number that neither party chooses: a value from a public
randomness beacon, or any pre-announced future public quantity. It has to be
outside the agency's control as well as the system's, or an agency can
target the system it dislikes.

Third, the sample follows by arithmetic. Each line identifier is combined
with the published number by a standard keyed function, the list is put in
the order that produces, and the first $m$ lines are the sample. The order
cannot be anticipated before the number exists and can be reproduced by
anyone afterwards, so the system, the agency and any third party arrive at
the same sample from two published values.

The system then does all of the field work, unchanged. The agency draws
nothing, holds no device and sends no crew: its entire burden is publishing
one number per cycle. What the system gets in return is the ability to prove
it sampled fairly, which at present it cannot do, because the draw happens
inside its own spreadsheet.

\textbf{Substitution is a free re-roll.} Where access fails the rule says to
select another line at random, which, run by the system, turns an address it
would rather not open into an address it did not open. Under a published
seed the replacements are fixed in advance, so a refused address has one
predetermined successor.

\textbf{The pool is the system's residue rather than the column.}
\S141.84(b)(5) excludes lines already confirmed non-lead by a two-point
visual inspection, so a system chooses what it will be tested on by choosing
what to inspect first. A rate $r$ measured on a share $c$ of the column
bounds the column at $[cr,\; cr + 1 - c]$ and no better, and the exclusion is
not neutral: an exterior inspection is precisely the method that cannot see
a lead lining, so the excluded stratum is enriched rather than depleted in
what the audit is looking for.

That last point is the sharper one, and it is not really about coverage. An
audit records $d\,\ell$, the product of the method's sensitivity and the
truth. Two-point visual inspection of a pipe exterior has $d \approx 0$ for a
lining on the inside, so a validation carried out perfectly, at the full
sample the rule demands, returns zero whatever is in the ground. The column
is not unaudited. It is audited by an instrument that cannot see the one
material that hides there, which is why the remedy is an instrument rather
than a new obligation.\cite{DeSantis2026}

\subsection{What strategy-proofness does and does not cover}

Theorem~3 is a statement about $\mathbb{E}[\hat T]$: no filing strategy
moves the expected need estimate. The allocator pays
$a(\hat T) = \max(\min(F_0, c\hat T),\, \theta^{*}\hat T)$, which is not
linear in $\hat T$, so $\mathbb{E}[a(\hat T)] \neq a(\mathbb{E}[\hat T])$
and a state that cannot move the mean can still move the payment by moving
the variance. Strategy-proofness in expectation is not strategy-proofness
in payment, and we price the difference rather than assert it away.

The payment map has two kinks with opposite curvature. Near the
proportional threshold it is convex, so audit noise raises expected
payment; near the need-cap boundary it is concave, so noise lowers it.
The lever is the realized sample, because a line that cannot be read in
the field does not enter the audit, and the field non-response rate is
unmeasured (Supplementary Note~4). Rather than assume a rate we take the
limit: every audit driven to the thirty-line statutory minimum, which is
the most obstruction can buy. The gain is monotone in the fraction of
lines lost, so nothing short of that pays more.

At that bound nine states gain and four lose. The gains total
\$19,091,015, **0.71\% of one year's pool**, and \$17,969,589 if every
gaining state degrades at once. Against the \$426.6 million of
year-to-year swing the damper removes, the \$271.1 million the repair
moves between states, and the \$33.3 million the audit itself costs, the
channel is small; it is not nothing, and for three states it is worth more
than a tenth of their own allotment (Supplementary Note~2).

\textbf{The audit therefore samples to a quota.} Drawing a fixed sample and
keeping whatever resolves lets non-response buy variance; drawing until the
planned number of lines has been \emph{resolved} does not, because the
realized sample is then the planned one whatever the response rate. The
channel closes outright, and non-response becomes a cost rather than an
incentive: the plan grows as $1/(1-f)$ in the field non-response rate.
That rate is unmeasured (Supplementary Note~4), so what the rule costs is
not known --- but what it is worth is, and the two cross at $f = 36.5\%$.
Below that the quota rule is cheaper than the channel it closes: at
$f = 10\%$ it adds \$3.7 million against \$19.1 million of exposure, and at
$f = 20\%$, \$8.3 million. A third of a sampled line list being unreadable
is the number that would change the recommendation, which makes it a
question with an answer rather than a caveat.

\subsection{The estimator and its parameters}

$\hat r_i$ is the raw mean of the state's audit, unshrunk, which is what
makes the strategy-proofness identity exact rather than approximate for
the audited system; until the audits exist, each state's measured
resolution flow stands in for its audit, and the pooled lines-weighted
yield ($r_0 = 0.117$) stands in only for states with no measurement at
all; both are labeled interim. The silence prior $\hat\pi_0 = 0.050$ is
the national expected lead per connection, computed from $r_0$, so the two
are not independent inputs. $\hat v_i$ is the same construction on the
declared-lead pool, and its interim value is $v_0 = 1$, because no
verification audit has been drawn and one is exactly what the current
rule assumes about that column. That interim is worth one more sentence,
because of where it bites: under the cap a declared line is worth the full
$c$ rather than $\theta^{*}(1-s_i)$, some thirty times more, and padding the
capped state's column to the cap boundary would recover \$15,632,842 --- the
same number, identically, as the freed-from-floor figure above, both being
$F_0 - c\hat T$ for that state. The verification audit is what separates the
two readings of it. Three pools are the LCRI inventory's,
not the design's: the identity holds for a report partitioned into any
number of declared classes, provided every credited class is audited at
its own rate and the uncredited ones are those a recipient gains nothing
by using. A program separating lead from galvanized lines requiring
replacement --- as some state revolving funds do --- applies the same
estimator with four terms rather than three. Every allotment in
Table~\ref{tbl:calibration} is therefore unchanged by the third rate: the
estimator differs from face-value crediting in what it can be held to,
not yet in what it pays, and strategy-proofness against over-declaration
is a property of the audited system exactly as it is for the other two
rates. A verified line is priced at the same \$524 for both sides as a
resolved one, because a physical verification costs the same whatever the
material: the pipe is found by excavating on either side of the meter pit
or curb stop, and there is no cheaper place to look for a line already
called lead (Supplementary Note~4).
Audit sizes are the largest of the applicable criteria, floored at 30
lines and capped at the pool sampled: one standard deviation of sampling
error, priced at the state's marginal dollars per line, below
$\varepsilon A$ with $\varepsilon$ of 0.1 percentage points; and enough
lines that the
estimate lands on the same side of the nearest regime boundary as the truth with
$z = 2$ standard deviations to spare. Propagating the sampling error through the allocation rather than
reasoning about it is what a National Academies panel recommended for
formula programs and left program-specific.\cite{NRC2003} Both are
evaluated with the finite-population correction and at each pool's own
anticipated rate,
smoothed to the Agresti--Coull point\cite{AgrestiCoull1998}
$(m r + 2)/(m + 4)$ so that a pool with no lead found so far is still
charged for the rate a zero count is consistent
with.\cite{Hanley1983} That gives
63,510 lines nationally, costed at published records-review and
field-verification rates;\cite{AWWA2022costs} Supplementary Table~2 gives
each state's own. Two ranges
are computed and reported separately, both by Monte Carlo through the
water-filling solution so that a state's range carries every other state's
error as well as its own. The pre-audit range draws each state's true
yield from the distribution its own resolutions support, on the Jeffreys
prior for a rate,\cite{BrownCai2001} widened to the companion's
cluster-bootstrap interval by matching its standard deviation, then draws the audit from the pool that results; the audit range holds the yield at its calibrated value and draws
only the audit. The first is what the evidence leaves open, the second is
what the sizing rule promises. Both are needed: ten of the twenty
measured states resolved \emph{negative} net lead over the panel and are
credited at the clipped zero, a pool held at a measured zero returns zero
from every audit, and the audit range alone would report certainty the
evidence cannot support. 
The connection anchor is a mis-statement risk rather than a strategic one:
$\partial \hat T/\partial K = \hat\pi \approx 0.05$, a ten per cent error
in every state's count at once moves 0.8\% of the pool and at most 0.21
percentage points of any one state's share, and because the unfiled stock
is floored at zero, understating $K$ lowers the understating state's own
payment.

\subsection{Allocation, damping and validation}

Equation~\eqref{eq:allocator} is solved by raising a single price per
expected line until the pool is exactly spent, states dropping out to
their floors as it rises --- a water-filling fixed point, and nothing to do
with water. Conservation is exact by construction and the fixed point is
unique whenever floors do not exhaust the pool. The damping gain is the steady state of the
local-level Kalman recursion, $\lambda = (\sqrt{q^2+4q}-q)/2$, with the
signal-to-noise ratio $q$ estimated from the panel's own revisions: the
FY2025 revision (a data-source change under an unchanged statute)
identifies revision noise and the FY2026 revision identifies drift,
giving $\lambda = 0.046$; audited baseline resets bypass the damper.
Reset invariance is verified by Monte Carlo in which the state resolves
its most lead-likely unknowns first with perfect foresight. The replacement cost is swept across the costing study's published
range and the per-line costs cities report, because the need cap binds
only below $F_0/c$ and so depends on a price the formula does not
set.\cite{Chicago2026cost} All computations are deterministic under fixed seeds; the full system is
implemented with 100\% branch test coverage, including a battery that
recomputes every table cell, every figure and every quoted value in
the pin tables from the raw panel on each run.

\subsection{Data availability}

The manuscript, the assembled panel, the diagnostics and the archived
source documents are deposited as one record
(DOI 10.5281/zenodo.22552146; the concept DOI 10.5281/zenodo.21808384
resolves to the latest version). The quarterly inventory vintages that
no longer exist upstream are included with checksummed provenance.

\subsection{Code availability}

All analysis code, the allocation module and the tests are in that
record and at \url{https://github.com/Gustolandia/documented-need}
(tag v1.6.0), under the MIT licence.

\subsection{Acknowledgements}

We thank Susan Masten for reading the manuscript and for the judgement
that its conclusions should not be softened for being strong, and the
state primacy-agency staff who answered questions about reporting
channels.

\subsection{Author contributions}

G.P.R.: conceptualization, methodology, formal analysis, software,
derivation of the mechanism, writing --- original draft. K.C.E.:
validation, data-quality assessment and industry-practice review from the
utility side, writing --- review and editing.

\subsection{Competing interests}

K.C.E. is employed by CDM Smith, an engineering firm active in lead
service line replacement; the firm had no role in study design, analysis,
or the decision to publish. G.P.R. declares no competing interests.

\bibliographystyle{unsrtnat}
\bibliography{references}

\end{document}

%% file: nw_table1.tex
\begin{tabular}{lrrrrl}
\toprule
State & $\hat T$ audited & Current & Prior & Audited (pre-audit range) & Regime \\
\midrule
Illinois & 975,980 & \$295.6M & \$365.5M & \$396.2M \ci{307.5}{465.9} & proportional \\
Michigan & 269,937 & \$143.5M & \$101.5M & \$109.6M \ci{92.1}{123.0} & proportional \\
Florida & 262,587 & \$32.8M & \$106.0M & \$106.6M \ci{27.9}{274.1} & proportional \\
Texas & 244,503 & \$76.6M & \$199.2M & \$99.3M \ci{82.9}{123.3} & proportional \\
Pennsylvania & 162,904 & \$126.1M & \$63.8M & \$66.1M \ci{48.1}{116.9} & proportional \\
New Hampshire & 38,716 & \$27.5M & \$27.5M & \$27.5M \ci{27.5}{31.5} & floor \\
Oregon & 3,228 & \$27.5M & \$27.5M & \$27.5M \ci{0.1}{27.5} & floor \\
Utah & 946 & \$27.5M & \$27.5M & \$11.8M \ci{11.2}{27.5} & need-capped \\
\bottomrule
\end{tabular}